\documentclass[aps,prd,twocolumn,showpacs]{revtex4}
\usepackage{graphicx}

\begin{document}

\newcommand{\be}{\begin{equation}}
\newcommand{\ee}{\end{equation}}
\newcommand{\e}{\epsilon}
\renewcommand{\d}{\delta}
\newcommand{\ck}{\hat{\chi}}
\newcommand{\dk}{\hat{\delta}}
\newcommand{\rhob}{\bar{\rho}}
\newcommand{\x}{\vec{x}}
\renewcommand{\k}{\vec{k}}
\newcommand{\g}{\gamma}
\renewcommand{\a}{\lambda}
\renewcommand{\l}{\lambda}
\newcommand{\pps}{\mathcal{P}}
\newcommand{\Like}{\mathcal{L}}

\title{Smoothing spline primordial power spectrum reconstruction}
\author{Carolyn Sealfon}
\email{csealfon@physics.upenn.edu}
\author{Licia Verde}
\email{lverde@physics.upenn.edu}
\author{Raul Jimenez}
\email{raulj@physics.upenn.edu}
\affiliation{
Dept. of Physics and Astronomy, University of Pennsylvania, Philadelphia,
PA 19104, USA.}

\begin{abstract}

We reconstruct the shape of the primordial power spectrum (PPS) using
a smoothing spline.  Our adapted smoothing spline technique provides a
complementary method to existing efforts to search for smooth features
in the PPS, such as a running spectral index.  With this technique we
find no significant indication with WMAP first-year data that the PPS
deviates from Harrison-Zeldovich and no evidence for loss of power on
large scales.  We also examine the effect on the cosmological
parameters of the additional PPS freedom.  Smooth variations in the
PPS are not significantly degenerate with other cosmological
parameters, but the spline reconstruction greatly increases the errors
on the optical depth and baryon fraction.
\end{abstract}


\maketitle

\section{Introduction}
The standard $\Lambda$CDM cosmological model with a Harrison-Zeldovich
primordial spectrum successfully predicts a wide range of current
observations (e.g.\cite{Spergel03}), implying that just five
cosmological parameters may describe the universe's composition and
evolution.
Current data can begin to probe the universe's primordial
inhomogeneities.  A power-law primordial power spectrum fits the
cosmic microwave background (CMB) and large-scale structure
observations very well \cite{Spergel03}.  There remains, however, the
tantalizing possibility that there could be structure in the
primordial power spectrum (PPS).  Theories of the universe's birth,
such as inflation or the ekpyrotic model, predict specific primordial
power spectra \cite{Easther04, Khoury03}; different models of
inflation predict different deviations from a pure power law.  Thus, the
statistical significance of scale-dependent features in the PPS can
support and/or rule out models of the very early universe and test the
inflationary paradigm.

While we cannot directly measure the PPS, the first-year WMAP
observations \cite{Hinshaw03} of the cosmic microwave background (CMB)
provide detailed information about the matter power spectrum as far
back in time as we can see.  The CMB power spectra depend on the PPS
through a well-known transfer function.
Large-scale structure data can also
probe the PPS. At present, most of the large-scale structure statistical
power is on scales smaller than those probed by CMB experiments, and such
analysis is complicated by non-linear growth of structure, bias, and
redshift space distortions.  Here, we do not consider large-scale structure,
and focus instead
on the statistical significance of features detected in the
PPS on CMB scales, such as a running of the spectral index and a
decrease in large-scale power, e.g. \cite{Peiris03, Hannestad03,
MukWang03c}.  We use a nonparametric approach that does not assume any
particular PPS model, but merely a smooth PPS function.

Typically, when fitting a model to data, one simply finds the
parameters that minimize the error.  When recovering a function from
discrete data (such as the CMB angular power spectra), however, there
are potentially infinite degrees of freedom and finite data.  There is
no limit on the number of parameters describing the function, so it is
possible to choose a function that interpolates the data to make the
error nearly zero.  Since the data is noisy, clearly such an
interpolating function will have features created by the noise that do
not exist in the true underlying function.  Yet if one uses too few
parameters (or the wrong choice of parameters), the fitting function
could miss features that are actually there.  In nonparametric
inference, the data themselves can determine just how many effective
degrees of freedom to give a function to recover the signal without
``fitting the noise''.  For background on nonparametric inference in
astrophysics, see \cite{Wasserman01}.

Some of the previous work to probe the shape of the PPS relies on
parametric models predicted by inflation \cite{Peiris03}, piecewise
linear reconstructions \cite{Bridle03, Blanchard03}, or linearly
interpolating between sliding bins \cite{Hannestad03}.  To invert the
transfer function convolution and directly re-create the PPS is quite
challenging \cite{Kogo03, Shafieloo03, TVDS04}, as computational
limitations force cut-offs or sampling that cause wiggles in the
reconstructed PPS.  Wavelets provide a rigorous nonparametric method
to search for sharp features as well as trends in the PPS
\cite{MukWang05, MukWang03b, MukWang03c}, but how to choose the number
of wavelets to use is a subjective choice.  Principle component
analysis of the CMB can detect important departures from a
scale-invariant PPS \cite{Leach05, Tak03}, but the PPS model used can bias
the recovered cosmological parameters.  Clearly, multiple methods are
needed to cross-check each other and contribute their respective
strengths to our understanding.

We explore a powerful nonparametric method to test the statistical
significance of smooth deviations from a scale-invariant, flat PPS.
The technique we use, complementary to the above methods, is a
smoothing spline.  Smoothing splines are a well-developed method used
frequently in nonparametric function reconstruction
\cite{GreenSilverman}, but this is their first application to the PPS.

As a side benefit, our analysis allows us to investigate degeneracies
between the PPS shape and the cosmological parameter values recovered
from CMB data.  There is concern (e.g., \cite{Blanchard03}) that
structure in the PPS could change WMAP's cosmological parameter
results, possibly even eliminating dark energy.  By allowing the PPS
function more degrees of freedom, we can test the robustness of the
cosmological parameters to flexibility in the PPS.

In the sections below, we explain the smoothing spline method and how
we apply it to reconstruct the PPS.  We then demonstrate the
application of our method on mock data sets, and present reconstructed
primordial power spectra from the WMAP first-year data.  We compare
the smoothing spline model-testing results with the Akaike and
Bayesian information criteria \cite{AIC, BIC}.  Lastly, we explore how
increasing the flexibility in the PPS affects the recovered
cosmological parameters and their errors.

\section{Smoothing spline}
\label{section:spline}

Here we review the smoothing spline technique (see, e.g.,
\cite{GreenSilverman}).  

Suppose one wants to recover a function $f(x)$ based on measurements
of $f$, denoted $\hat{f}$, at a discrete set of points $x_i$.  In the
smoothing spline method, one adds a roughness penalty to the
chi-square error, and minimizes their sum:
\be
\label{eq:S}
S(f) = \sum_{i=1}^n \frac{(\hat{f}(x_i)-f(x_i))^2}{\sigma_i^2} + \a \int_{x_1}^{x_n} (f''(x))^2 dx .
\ee
The $x_i$ are called ``knots'', and $\sigma_i$ denotes the
standard deviation at each point.

The integral of the function's second derivative squared is called the
``roughness penalty'' and disfavors jagged and bumpy functions that
``fit the noise''.  If $f(x)$ is a general function with infinite
degrees of freedom, minimizing the error alone will recover features
created by random noise in the data, as well as the smooth underlying
function's features.  The roughness penalty effectively reduces the
degrees of freedom to only allow smooth functions.  The smoothing
parameter, $\a$, determines the amount of roughness allowed; larger
$\a$ implies a smoother function.  As $\a$ goes to infinity, the
problem becomes one of linear regression, and as $\a$ goes to zero,
the problem becomes one of interpolation (``connect the dots'').
Clearly, a method is required for choosing $\a$, since different $\a$
give different answers for $f$.  Cross validation, discussed later,
provides a rigorous statistical technique for choosing the optimum
smoothing parameter.

Notice that the first term in $S$ depends only on the values of $f$ at
the points of measurement.  It can be shown (e.g.,
\cite{GreenSilverman}) that the function that minimizes the roughness
penalty for fixed values of $f(x_i)$ is a cubic spline: an
interpolation of points via a continuous piecewise cubic function,
with continuous first and second derivatives.  The continuity
requirements uniquely determine the interpolating spline, except at
the boundaries.  Here, we use the assumption-independent boundary
condition that the third derivative also be continuous at the first
and last interior knots.

\section{Applying the smoothing spline to the primordial power spectrum}

The smoothing spline is a natural technique to reconstruct the PPS,
since we expect it to be a smooth function.  As we do not have direct
measurements of the PPS nor unlimited computing power, we cannot bring
the full power of the smoothing spline method 
to bear on this problem.  Nonetheless, it is an
informative tool that can only grow more illuminating with better data
and computing power.  The method allows us to give the PPS much more
freedom than parametric models, and measure how the data take advantage of
this extra parameter space.

The data we use to recover the PPS are the first-year WMAP
measurements of the CMB temperature (TT) and temperature-polarization
cross-correlation (TE) angular power spectra as a function
of multipole moment $\ell$ \cite{Hinshaw03, Kogut03}.  
Large-scale structure data could provide more information, and 
while the largest scales probed by large-scale structure now
overlap with the small scales probed by CMB, we leave an analysis that
includes galaxy surveys data for future work. Here, we concentrate on
developing the method to be applied in the cosmological context to
reconstruct the primordial power spectrum, where there is a complicated
non-linear function connecting the PPS to the observed angular power
spectrum of the CMB. In addition, we are interested in features on scales
probed by CMB data, about which there has been controversy in the
literature \cite{Peiris03, Hannestad03, MukWang03c, Bridle03, Blanchard03, Kogo03, Shafieloo03, TVDS04, MukWang05, MukWang03b}.
We write the primordial spectrum in terms of the scalar metric
fluctuation, $\chi$, as

\be 
(2 \pi)^3 P(\k,t) \delta^D(\k +\k')=\langle \ck(\k,t)\ck(\k',t) \rangle,
\ee 
where $\langle \rangle$ denote the ensemble average.  The
dimensionless PPS in an isotropic universe is then, 
\be
\pps(k)=\frac{1}{2 \pi^2} P(k) k^3 .  
\ee 

We generalize the smoothing spline method, described in the previous section,
to apply to the PPS.  In place of the chi-square error, we use the log
likelihood function
as computed by the code provided by the WMAP team
\cite{Verde03}.  The theory $C_\ell$ depend on $\pps(k)$ and
cosmological parameters via the transfer function, $C_\ell=T(\pps(k))$,
computed using the CMBFAST code \cite{SZ96} modified to
accept a spline PPS. We place the roughness penalty on $\pps(k)$ as
a function of $\ln(k)$ because we wish to discern deviations from a
power-law PPS.

The penalized error we minimize (compare to Eq.(\ref{eq:S})) is thus,

\be
S(\pps) = -\ln \Like(\pps, \mathrm{cosmological \, parameters})+ \a F(\pps), 
\label{eq:Spps}
\ee 
with the roughness penalty given by,
\be
F(\pps)= \int_{\ln k_\mathrm{min}}^{\ln k_\mathrm{max}} (\pps''(\ln k))^2 d\ln k.
\ee 

Since the transfer function is a convolution over $\pps(k)$, we
cannot choose the knots so each knot corresponds univocally to a data
point.  
Due to computational limitations, we cannot choose nearly as many knots as
data points.  
Thus, we are only sensitive to smooth, large-scale
features and the general shape of the primordial power spectrum.
However, if there are sharp features that are sufficiently
statistically significant, this method may  still detect a deviation
from scale-invariance.  We space the knots equally in $\ln(k)$ between
$k_\mathrm{min}=1 \times 10^{-5}$ and $k_\mathrm{max}=0.2$
\footnote{A more rigorous analysis could let the data choose the
placement of knots using statistical evidence.  This will be the
approach for future work.}.  
Note that there is a trade-off between the number of knots and the
$\a$ values:  reducing the
number of knots forces the function to be smoother, and thus has
a similar effect to increasing $\a$ (and vice-versa).

The smoothing spline method thus allows for broad variations in the
PPS.  It is general enough to include shapes produced by a running of
the spectral index or a gentle cutoff at low multipoles.  The
smoothing parameter determines the above-mentioned balance to
recover signal without fitting noise.

\subsection{Markov Chain Monte Carlo}

For the smoothing spline PPS reconstruction, we assume a flat universe
and treat as free parameters the physical matter density $\Omega_m
h^2$, the physical baryon density $\Omega_b h^2$, the sound horizon's
angular size $\cal{A} $, and the optical depth $\tau$ (through the
parameter $Z=e^{-2 \tau}$), e.g., \cite{KMJ2002}.  In addition to these
four cosmological parameters, our parameter space includes the value
of $\pps(k)$ at each knot.  Thus, the parameter space is nine
dimensions for a five-knot spline, twenty-one dimensions for a
seventeen-knot spline.  We seek the point in parameter space that
minimizes Eq.(\ref{eq:Spps}).  It is computationally impossible to
test all combinations of the cosmological parameters and values of
$\pps(k)$ at the knots.  Therefore, we use the Markov Chain Monte
Carlo (MCMC) method to sample the parameter space, map the log
likelihood (or, in this case, $S$) surface as a function of parameter
space, find the extremum, and locate the one-sigma bounds.

MCMC evaluates and records the value of $S$ at discrete points in such
a way that the density of the recorded points in parameter space is
proportional to the posterior distribution.  As it evaluates each
point, it decides whether or not to add the coordinates of that point
and its $S$ value to a list.  This list is called a ``Markov chain''.
We use the Metropolis-Hastings algorithm to select which points to put
on the list and how many times each point is repeated. 
See \cite{Verde03}, section 3, or \cite{Gilks,cosmomc} for a more detailed
description.  MCMC thus maps the likelihood surface to as much
accuracy as needed.

The details of our implementation follow.
Our MCMC code computes and records $S$ and the
roughness penalty, and maximizes $-S$ rather than the log likelihood.
De Boor's code \cite{deBoor} calculates the cubic spline that interpolates
the knot values.  To choose the optimum step size for fastest
convergence, we find the covariance matrix for a preliminary,
short chain and take MCMC steps along the eigenvectors of the
covariance matrix with eigenvalue step sizes.  To check convergence
and mixing, we use cusum path plots \cite{YuMykland} calibrated from
and compared with a converged WMAP chain \cite{Verde03}.

Using importance sampling (re-weighting the Markov chain)
\cite{Gilks,cosmomc}, we can vary $\a$ slightly without re-running a given
chain.  Since we store the value for $F$ as well as $S$ for each point
in the chain, we can re-weight each point's contribution to the
posterior density by a factor of $e^{(\a_o-\a)F}$, where $\a_o$ is the
smoothing parameter for the initial chain, so the resulting density
corresponds to a smoothing parameter $\a$.  This procedure only works
when $e^{(\a_o-\a)F} \ll e^{-S}$.  If the shape of $e^{-S}$ as a
function of parameter space changes significantly, the original Markov
chain will not have adequately explored regions of lowest $S$ for the
new $\a$, and importance sampling does not give meaningful results.

\subsection{Cross validation}

The smoothing spline requires an objective method to choose the
optimum value of $\a$ to recover the underlying function.
Cross-validation (CV) (see, e.g., \cite{GreenSilverman}) quantifies
the notion that if we have found the correct underlying function, the
result should be insensitive to more data.  The most rigorous
form of CV is referred to as ``leave out one'': run the analysis
leaving out one data point, get the optimum function, and compute
the error between that data point and the recovered function's value
at that point.  Then do this for each data point, sum the
resulting errors, and choose the $\a$ that minimizes this sum.

One-fold CV is too computationally expensive for us.  We have data
points at 899 $\ell$'s, so we would have to run 899 Markov chains if
we left each $\ell$ out once.  Because the transfer function lies
between the PPS and the data, we cannot apply the calculation shortcut
in \cite{GreenSilverman}.  Instead, we use $\frac{N}{2}$-fold cross
validation, where $N=899$ is the number of data points ($\ell$'s).  We
run a Markov chain using only half the data (every-other $\ell$) with
some initial $\a$.  We then take the coordinates of every point from
that Markov chain ($\pps$ knots and cosmological parameters), and
compute the log likelihood of the corresponding $C_\ell$ with respect
to the other half of the data, $\ln \Like_\mathrm{CV}$, for each
point.  Next we repeat this process with the two halves of the data
switching roles.  Using importance sampling, we choose the best-fit
$\pps$ for each half of the data for any nearby $\a$, and note the
corresponding $\ln \Like_\mathrm{CV}$ with respect to the other half.
Summing these two $\ln \Like_\mathrm{CV}$ gives the ``CV score'' for a
given $\a$.  The best $\a$ is the one with the highest CV score.

Since the roughness penalty is independent of cosmological parameters,
we keep them fixed for the CV Markov chains and vary only the PPS
\footnote{We split the CMBFAST code between the part of CMBFAST that
computes the transfer function and the part that performs the
convolution with the PPS, to make the chains run faster.}.  We check
how much fixing the cosmological parameters affects the CV results by
running CV for two different sets of fixed cosmological parameters,
and once allowing the cosmological parameters to vary.

An important assumption of CV is that the data ``left out'' to perform
CV is not correlated with the remaining data.  If the data are
correlated, the smoothing parameter will be under-estimated by CV, and
a generalization of CV for correlated errors should be used
\cite{Opsomer01}. When performing $\frac{N}{2}$-fold CV with full-sky CMB
experiments such as WMAP, the omitted data points are not strongly
correlated to the remaining data points.  In fact, for the WMAP data,
neighboring points are negligibly correlated (see figure (14) of
\cite{Hinshaw03}).

\section{Testing the Smoothing Spline Method on Mock Data}

To assess the performance of our method, we test it on mock data sets
created from select primordial power spectra using the publicly
available HEALPix software
\footnote{{\tt http://www.eso.org/science/healpix/}}.  The power spectra we
choose to examine are A) the best-fit running of the spectral index
($\alpha_s=-0.047$) with the best-fit WMAP parameters, B) a running
one-sigma steeper than the WMAP best-fit ($\alpha_s=-0.087$), and C) a
scale-invariant power spectrum with a sharp cutoff at $k=0.002$ Mpc
\footnote{The cutoff is introduced by multiplying the flat spectrum by
$\frac{1-e^{-2(k/k_s)^4}}{1+e^{-2(k/k_s)^4}}$ with $k_s=0.002$.}.  We
choose these PPS because a running and a cutoff are two features that
have been claimed to be consistent with the data \cite{Peiris03,
Bridle03, Shafieloo03, Cline03, Contaldi03, Feng03a}, and we would like to
determine how well our method would detect them.

The $C_\ell$ errors are non-gaussian, especially at low $\ell$, and
the WMAP team uses a sum of a lognormal and a normal distribution to
approximate the non-gaussianity \cite{Verde03}.  Thus for case A we
reconstruct the spectrum twice: in one case using the WMAP likelihood
code, in the other case using the ideal likelihood (Eq.(4) of
\cite{Verde03}) for $\ell \leq 100$.  In all other cases we use the
WMAP likelihood code.  Our reconstruction uses a five-knot smoothing
spline and zero lambda.

\begin{figure}
\includegraphics{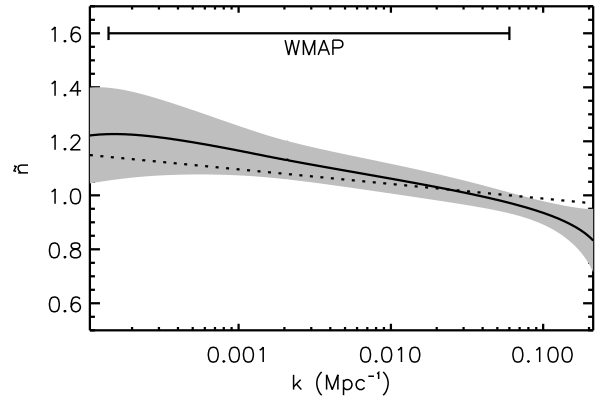}
\caption{The spectral index $\tilde{n}(k)$ of the reconstructed PPS from the
mock data with a running of $\alpha_s=-0.047$ and the
best-fit values for the WMAP running spectral index model for the rest
of the cosmological parameters.  The reconstruction used a smoothing
spline with five knots equally spaced in $\ln k$ and no roughness
penalty, and the likelihood approximation used by the WMAP team.  The
dark line denotes the mean, the shaded region denotes one standard
deviation, and the dashed line denotes the fiducial PPS.}
\label{fig:mock0}
\end{figure}

\begin{figure}
\includegraphics{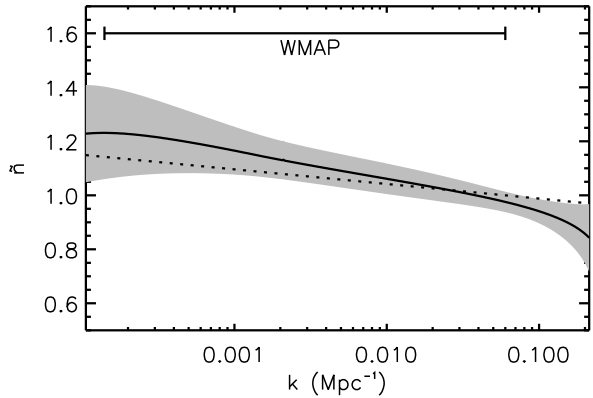}
\caption{The spectral index $\tilde{n}(k)$ of the reconstructed PPS from the
mock data with a running of $\alpha_s=-0.047$ and the
best-fit values for the WMAP running spectral index model for the rest
of the cosmological parameters.  The reconstruction used a smoothing
spline with five knots equally spaced in $\ln k$ and no roughness
penalty, and an exact likelihood calculation for $\ell=2-100$.  The
dark line denotes the mean, the shaded region denotes one standard
deviation, and the dashed line denotes the fiducial PPS.}
\label{fig:mock0b}
\end{figure}

\begin{figure}
                                                                                     
\includegraphics{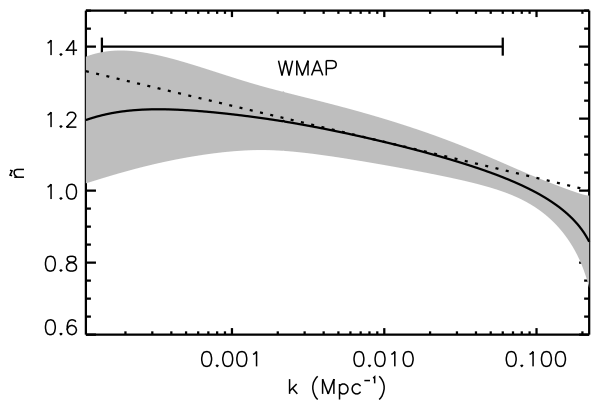}
\caption{The spectral index $\tilde{n}(k)$ of the reconstructed PPS from mock
data with a running of $\alpha_s=-0.087$ and the best-fit
values for the WMAP running spectral index model for the rest of the
cosmological parameters. The reconstruction used a smoothing spline
with five knots equally spaced in $\ln k$ and no roughness penalty.
The dark line denotes the mean, the shaded region denotes one standard
deviation, and the dashed line denotes the fiducial PPS.  The bend at
low $k$ is due to the spectrum hitting the prior.}
\label{fig:mock1}
\end{figure}

\begin{figure}
\includegraphics{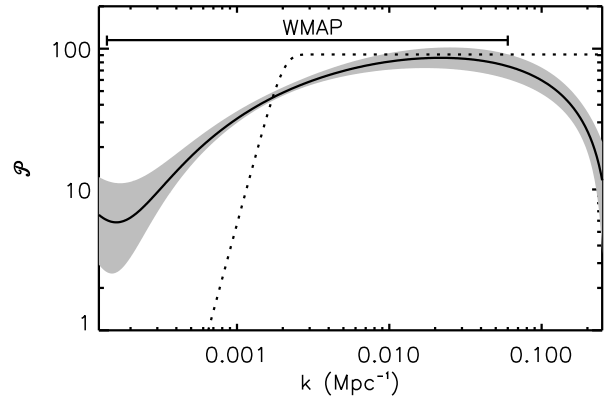}
\caption{The reconstructed $\pps(k)$ from mock
data with a cutoff at $k=0.002$ the best-fit
values for the WMAP flat PPS model for the rest of the
cosmological parameters. The reconstruction used a smoothing spline
with five knots equally spaced in $\ln k$ and no roughness penalty.
The dark line denotes the mean, the shaded region denotes one standard
deviation, and the dashed line denotes the fiducial PPS.}
\label{fig:mock3}
\end{figure}

We plot the spectral index as a function of $k$, $\tilde{n}(k)$ using
the definition \footnote{Note that $\tilde{n}(k)$ defined here differs
from $n(k)=1 + \frac{d\ln\pps}{d\ln k}$ defined in \cite{Spergel03}
Eq. (5) by a factor of two in the running, but the two conventions are
equivalent for constant $n$, and the running $\alpha_s$ is defined to
be their $dn_s/d\ln k$.  In other words, our $\tilde{n}(k)$ is the
exponent in their Eq. (3).},
\be
\tilde{n}(k)=1+\frac{\ln(\pps(k))-\ln(\pps(k_0))}{\ln k - \ln k_0},
\label{eq:n}
\ee
with $k_0=0.002$Mpc$^{-1}$ for cases A and B.
Case A is shown figures (\ref{fig:mock0}) and (\ref{fig:mock0b}),
where the former used the WMAP likelihood approximation, and the latter
used the ideal likelihood  calculation for $\ell \leq 100$ ignoring
noise and coupling due to sky cut.
Figure (\ref{fig:mock1}) shows the one-sigma running
(case B).  In all
the figures, the dashed line denotes the fiducial PPS, and the solid
line the mean reconstructed spectrum. 
Our method
successfully recovers the PPS over the WMAP-sensitive scales, and
detects a noticeable deviation from $\tilde{n}=1$.
The comparison of the two case A reconstructions, figures (\ref{fig:mock0})
and (\ref{fig:mock0b}), shows that the likelihood approximation does not
significantly affect the reconstructed spectrum.
The bend at low $k$ in figure (\ref{fig:mock1})
is due to the prior requiring $\pps(k)$ to be
greater than ten at the knots.  Clearly $\pps(k)$ must be positive,
but occasionally the interpolating spline between the knots goes
negative, and we must discard such steps in the Markov chain.
Choosing a lower prior of ten for the knot values helps keep the
interpolating spline from going negative.
The ``sharp cutoff'' PPS reconstruction, shown in figure
(\ref{fig:mock3}), hits a non-negative prior on the PPS knot values
more strongly, since splines with steep slopes are more likely to go
negative.  It resembles a running of the spectral index, showing how
our method tends to ``smooth out'' sharp features.

\section{PPS Reconstruction Results}

We next apply the smoothing spline reconstruction to the combination
 of the first-year WMAP TT and TE power spectra \cite{Kogut03,
 Hinshaw03}.  Our aim here is to test the statistical significance of
 possible smooth deviations from scale invariance; our ``null
 hypothesis'' is a scale-invariant ($n=1$) PPS.  Thus, as a benchmark,
 we run a Markov chain to find the best-fit parameters for such a
 model.  Since the volume of parameter space for wiggly power spectra
 is far greater than the volume for flat spectra, the Markov chain
 tends towards more curvy (higher $F$) spectra than flat ones, due to
 entropy.  One may compensate for this effect by increasing $\a$, but
 it is still very unlikely for a Markov chain started from a curved
 PPS to line up the spline's knots to retrieve a totally flat $n=1$
 spectrum.
To minimize burn-in and to make sure our chains include the null
hypothesis case, we start the smoothing spline Markov chains from a
scale-invariant PPS with cosmological parameters set to the $n=1$
model's best-fit parameters.

After running several preliminary chains with different numbers of
knots and different smoothing parameters, we settled with three
well-mixed, converged Markov chains with smoothing splines, one with
five knots and roughness penalty $\a=0.1$, one with five knots and
$\a=0$,and one with seventeen knots and roughness penalty $\a=2 \times
10^{-5}$.  
We use the roughness parameter of $2 \times 10^{-5}$ for the 17-knot
spline because it is very tiny constraint on the flexibility of the
PPS, but sufficient to allow the Markov chain to converge in a reasonable
amount of time.

\begin{figure}[ht]
\includegraphics{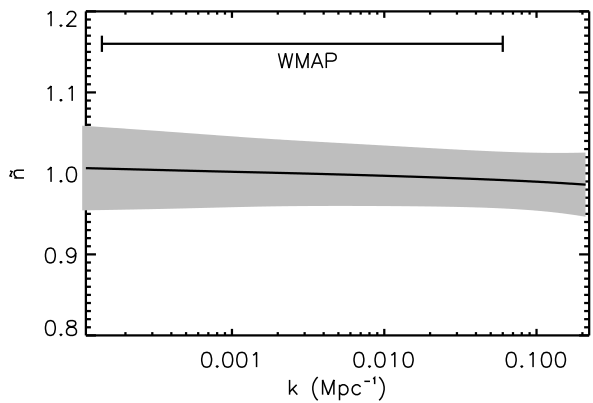}
\caption{The spectral index $\tilde{n}(k)$ of the PPS reconstructed using a
smoothing spline with five knots equally spaced in $\ln k$ and a
smoothing parameter of 0.1.  The dark line denotes the mean and the shaded
region denotes one standard deviation.}
\label{fig:nk5alpha}
\end{figure}

\begin{figure}[ht]
\includegraphics{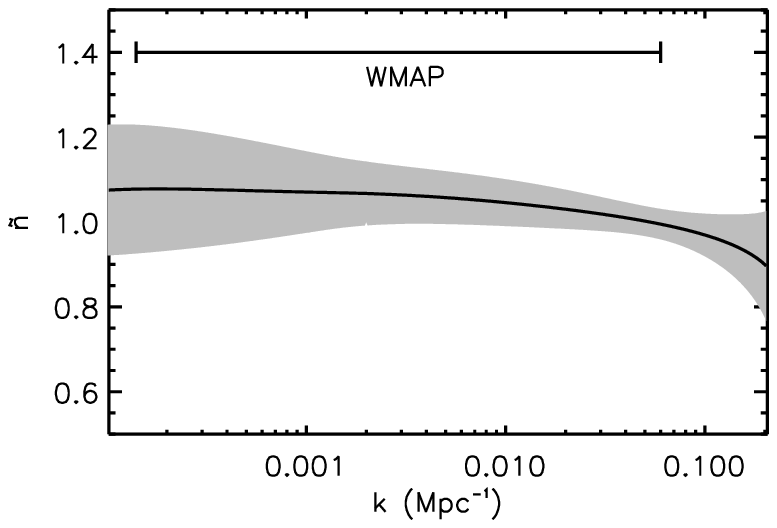}
\caption{The spectral index $\tilde{n}(k)$ of the PPS reconstructed using a
smoothing spline with five knots equally spaced in $\ln k$ and no
smoothing penalty.  The dark line denotes the mean and the shaded
region denotes one standard deviation.}
\label{fig:nk5}
\end{figure}

\begin{figure}[ht]
\includegraphics{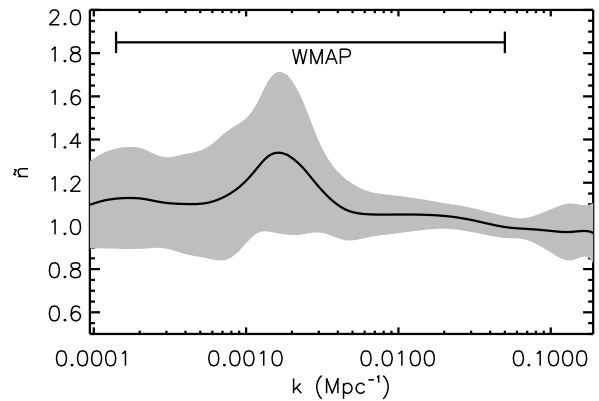}
\caption{The spectral index $\tilde{n}(k)$ of the PPS reconstructed using a
smoothing spline with 17 knots equally spaced in $\ln k$ and a
smoothing parameter of $2\times 10^{-5}$.  The dark line denotes the mean
and the shaded region denotes one standard deviation.}
\label{fig:nk17}
\end{figure}

Figure (\ref{fig:nk5alpha}) shows the mean spectral index
(Eq.(\ref{eq:n})) and one-sigma confidence interval for the WMAP data
for a smoothing spline with five knots and a smoothing parameter of
0.1.
The resulting PPS is indistinguishable from flat.  Removing the
roughness penalty yields the recovered PPS shown in figure
(\ref{fig:nk5}), which has a $\Delta\chi^2$ of only 1.79 away from
flat.  Giving the PPS spline a lot more freedom, seventeen knots and a
smoothing parameter of only $2 \times 10^{-5}$, one recovers the
function in figure (\ref{fig:nk17}).  Although this spline was given
much more parameter space and a much lower $\a$ than preferred by the
data, the one-sigma confidence interval is still consistent with a
flat $n=1$ spectrum.

In all these cases, the posterior probability density created by the
Markov chains concentrates in regions of low roughness penalty.  This
implies that the data do not require the additional flexibility of the
smoothing spline reconstruction.  For all the steps in all our runs
with finite $\a$, $\a F$ remains on the order of 0.1\% to 1\% of $\ln
\Like$. This means that moving far away from a flat power spectrum
cannot improve the likelihood sufficiently to overcome the roughness
penalty.

CV confirms this qualitative interpretation. We ran our
$\frac{N}{2}$-fold CV for two different sets of fixed cosmological
parameters, the best-fit flat and the five-knot no-lambda best fit, and
find an optimum $\a \gtrsim 0.1$ for a five-knot spline.  Therefore, the PPS
should be closer to scale invariant than the five-knot spline with
$\a=0$ results (figure \ref{fig:nk5}) and at least as close to scale
invariant as the five-knot spline with $\a=0.1$ (figure
\ref{fig:nk5alpha}).  Since increasing the number of knots or
decreasing $\a$ both increase the effective degrees of freedom, we can
also conclude that the PPS should be much closer to scale invariant
than the 17-knot spline with $\a=2 \times 10^{-5}$ result (figure
\ref{fig:nk17}).

Figure (\ref{fig:nk5}) shows a flatter PPS than (\ref{fig:mock0}),
with the same number of knots and same smoothing parameter ($\a=0$).
Thus, we conclude that if the PPS had a running of $|\alpha_s|\gtrsim
0.05$ the smoothing spline technique applied to WMAP data would have
recovered it.
The smoothing spline finds no evidence
for deviations from a scale invariant power spectrum from WMAP
first-year data.

\subsection{Comparing CV with other criteria}

The preference for a very flat PPS, indicated by the spline
reconstruction and its CV, can be cross-checked with alternative model-testing
techniques, such as Bayesian evidence.

We apply the Akaike information criterion (AIC) and Bayesian
information criterion (BIC) to compare the five-knot, $\a=0$ spline
with the scale invariant model.  These methods are improvements on the
standard likelihood ratio test \cite{Kendall} in that they rely on
fewer assumptions \cite{Liddle04}.  Both criteria add a penalty for
additional degrees of freedom to the negative log likelihood, so that
a lower value for this sum indicates the preferred model.  The AIC
\cite{AIC} comes from an approximation to the Kullback-Leiber
information entropy, used to measure the difference in goodness of fit
between two models.  The BIC \cite{BIC} approximates the Bayes factor,
or the ratio of the likelihood of two models with a flat initial
prior.  The BIC has a more stringent requirement to accept extra
parameters than the AIC \cite{Liddle04}.  Here we find that both
criteria prefer the simpler, scale-invariant model; see table
(\ref{table:ABIC}).  A difference in BIC of ten or more is considered
very strong evidence for the preferred model, and here we have a
difference of twenty-six favoring the flat spectrum over the five-knot
spline.  Even the weaker AIC prefers the simpler model by a difference
of six.

Our results are consistent with previous work applying Bayesian evidence to
determine the significance of the spectral index $n$ in the cosmological
parameter set (versus setting $n=1$).  Liddle \cite{Liddle04} applies
the AIC and BIC and finds that the data favor the simpler
scale-invariant model.  Trotta \cite{Trotta05} uses the Laplace and
Savage-Dickey approximations to the Bayes factor, and finds 2:1 odds
in favor of the Harrison-Zeldovich spectrum using CMB data.  Our tests
extend these results and compare the Harrison-Zeldovich model to
shapes with more flexibility than the one free parameter $n$.

\begin{table}
\begin{ruledtabular}
\begin{tabular}{ccc}
& 5-knot spline & minimal ($n$=1)\\  
\hline
AIC & 1448 & 1442\\
BIC & 1494 & 1468\\
\end{tabular}
\end{ruledtabular}
\caption{The Akaike information criterion and Bayesian information
criterion applied to the five-knot smoothing spline PPS with zero lambda
and the minimal cosmological model. Since preferred models have lower
AIC and BIC, both criteria prefer the minimal model.}
\label{table:ABIC}
\end{table}

\section{Sensitivity of Cosmological Parameters on the PPS}

\begin{table*}
\begin{ruledtabular}
\begin{tabular}{cccccc}
& 5-knot spline & 17-knot spline & power law & running & minimal ($n$=1)\\  
\hline
$\Omega_b h^2$ & 0.021$\pm$0.002 & 0.022$\pm$0.004 & 0.024$\pm$0.001& 0.023$\pm$0.002 & 0.0240$\pm$0.0004 \\
$\Omega_m h^2$ & 0.14$\pm$0.02 & 0.14$\pm$0.03 &  0.14$\pm$0.02 & 0.14$\pm$0.02 & 0.15$\pm$0.02\\
$h$ & 0.67$\pm$0.06 & 0.63$^{+0.18}_{-0.10}$ & 0.72$\pm$0.05 & 0.70$\pm$0.05 & 0.70$\pm$0.05\\
$\tau$ & 0.21$^{+0.08}_{-0.10}$ & 0.26$^*_{-0.08}$ & 0.166$^{+0.076}_{-0.071}$ & 0.20$\pm$0.07 & 0.18$\pm$0.05\\
\end{tabular}
\end{ruledtabular}
\caption{Cosmological  parameters and  errors  recovered by  smoothing
spline PPS's versus the WMAP  results \cite{Spergel03} for a power law
and running spectral index PPS and an $n$=1 PPS. The values and errors
for $\Omega_m$ remain quite  independent of PPS flexibility.  Allowing
smooth features in the PPS loosens the constraints on $\Omega_b$, $h$,
and $\tau$. * denotes hitting the prior.}
\label{table:CPs}
\end{table*}

Because the measured CMB power spectrum depends nonlinearly on both
the PPS and the cosmological parameters, there are degeneracies
between the PPS shape and the cosmological parameters.  We now
investigate how the increased freedom in the PPS allowed by the spline
technique affects the recovered values of the cosmological parameters.
Similar analyses yielding qualitatively similar results using multiple
CMB data sets, large-scale structure, and piecewise and band-power PPS
reconstructions, can be found in \cite{Bridle03, MukWang03c}.

\begin{figure*}[ht]
\includegraphics{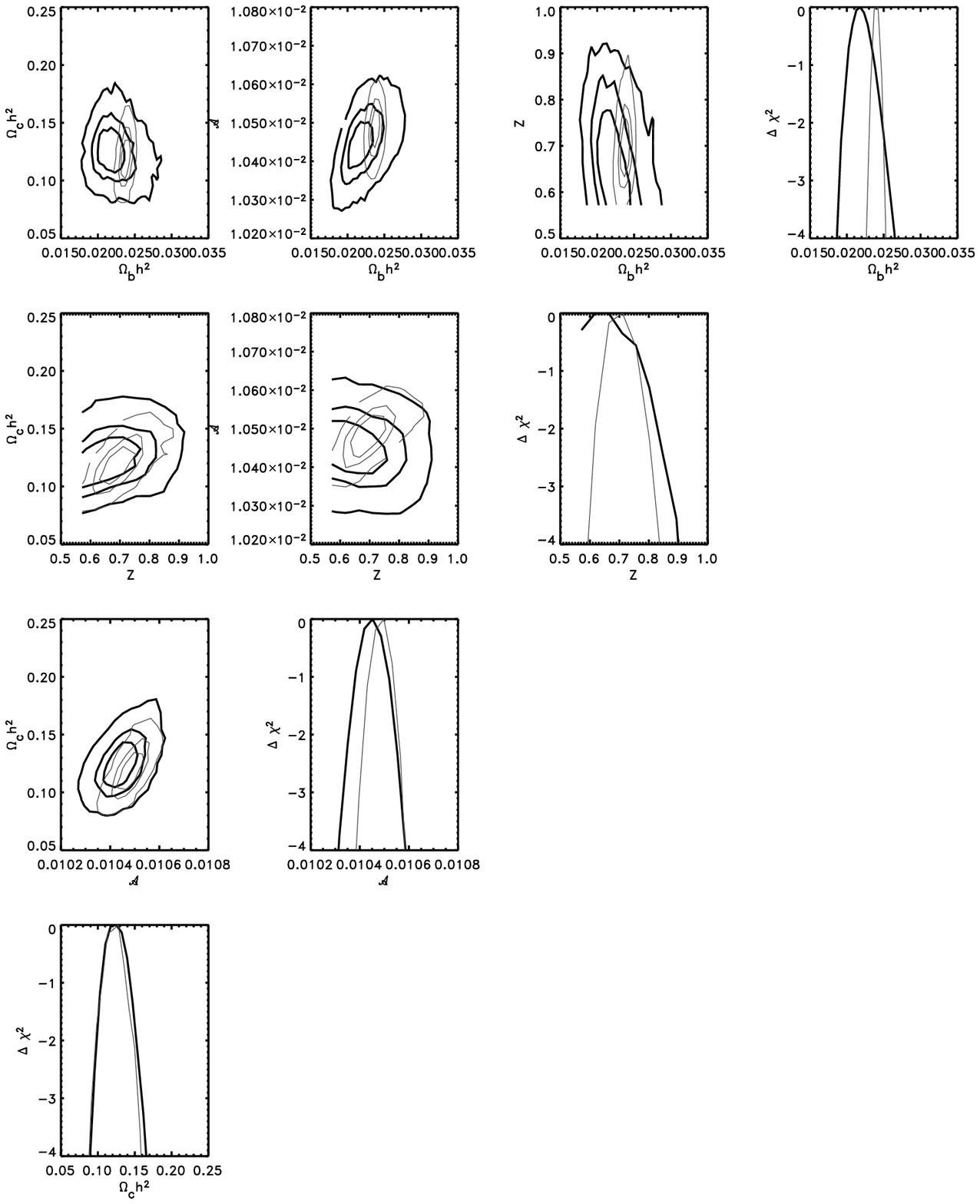}
\caption{Errors on cosmological parameters from a Markov chain run for
a smoothing spline $\pps$ with five knots and no roughness
penalty. Dark contours denote one sigma marginalized, one sigma joint,
and two sigma joint.  Over-plotted with thin, gray contours are the
errors for a $n=1$ flat PPS for comparison.  $\Omega_b$ is the baryon
fraction, $\Omega_c$ is the cold dark matter density, ${\cal A}$ is the
present angular size of the sound horizon at the last scattering
surface, and $Z=e^{-2 \tau}$ where $\tau$ is the optical depth.}
\label{fig:cp5}
\end{figure*}

\begin{figure*}[ht]
\includegraphics{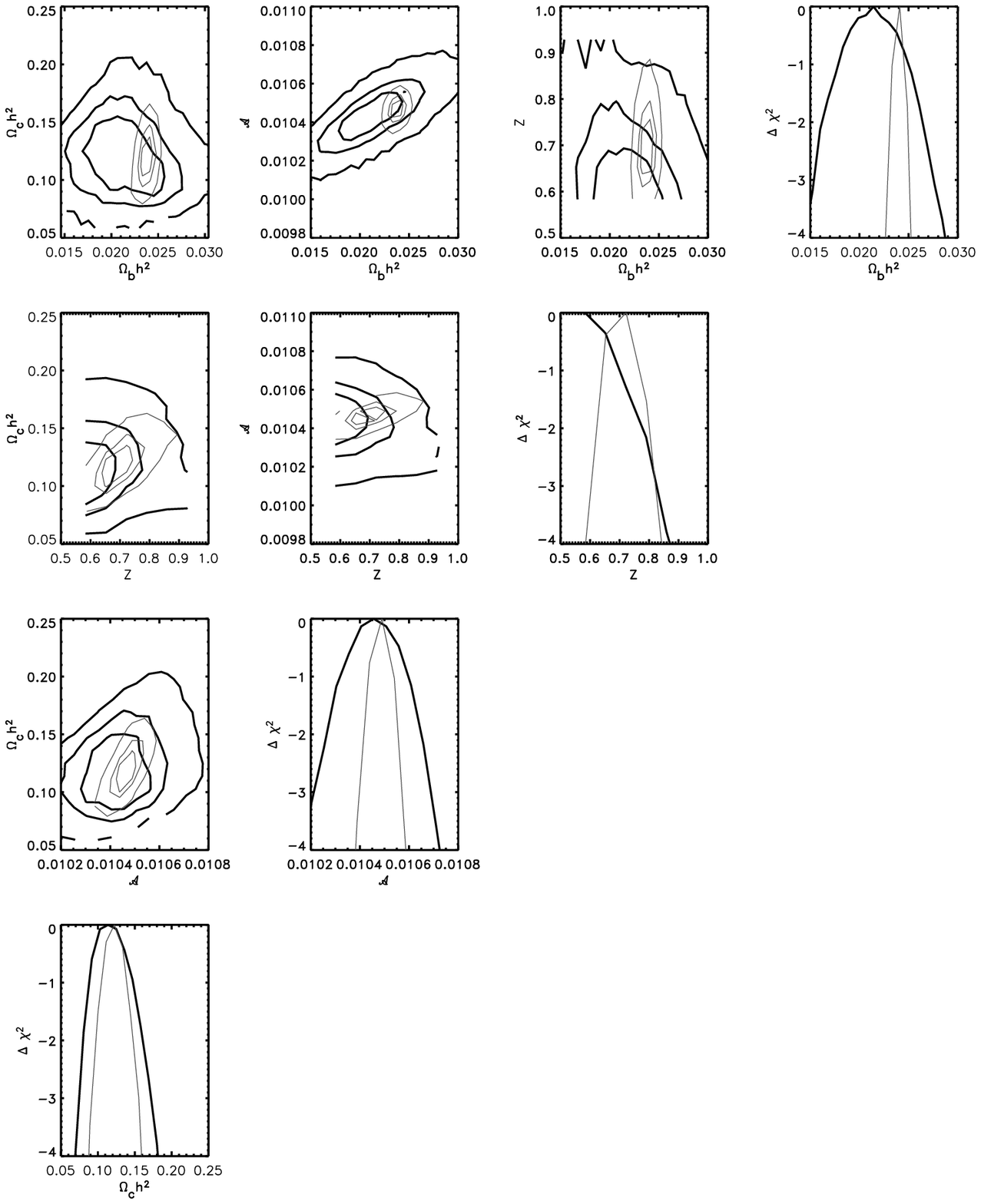}
\caption{Errors on cosmological parameters from a Markov chain run for
a smoothing spline $\pps$ with 17 knots and and a smoothing parameter
of 2x$10^{-5}$.  Dark contours denote one sigma marginalized, one
sigma joint, and two sigma joint. Over-plotted with thin, gray contours
are the errors for a $n=1$ flat PPS for comparison. $\Omega_b$ is the
baryon fraction, $\Omega_c$ is the cold dark matter density, ${\cal A}$ is
the present angular size of the sound horizon at the last scattering
surface, and $Z=e^{-2 \tau}$ where $\tau$ is the optical depth.}
\label{fig:cp17}
\end{figure*}

\begin{figure*}[ht]
\includegraphics{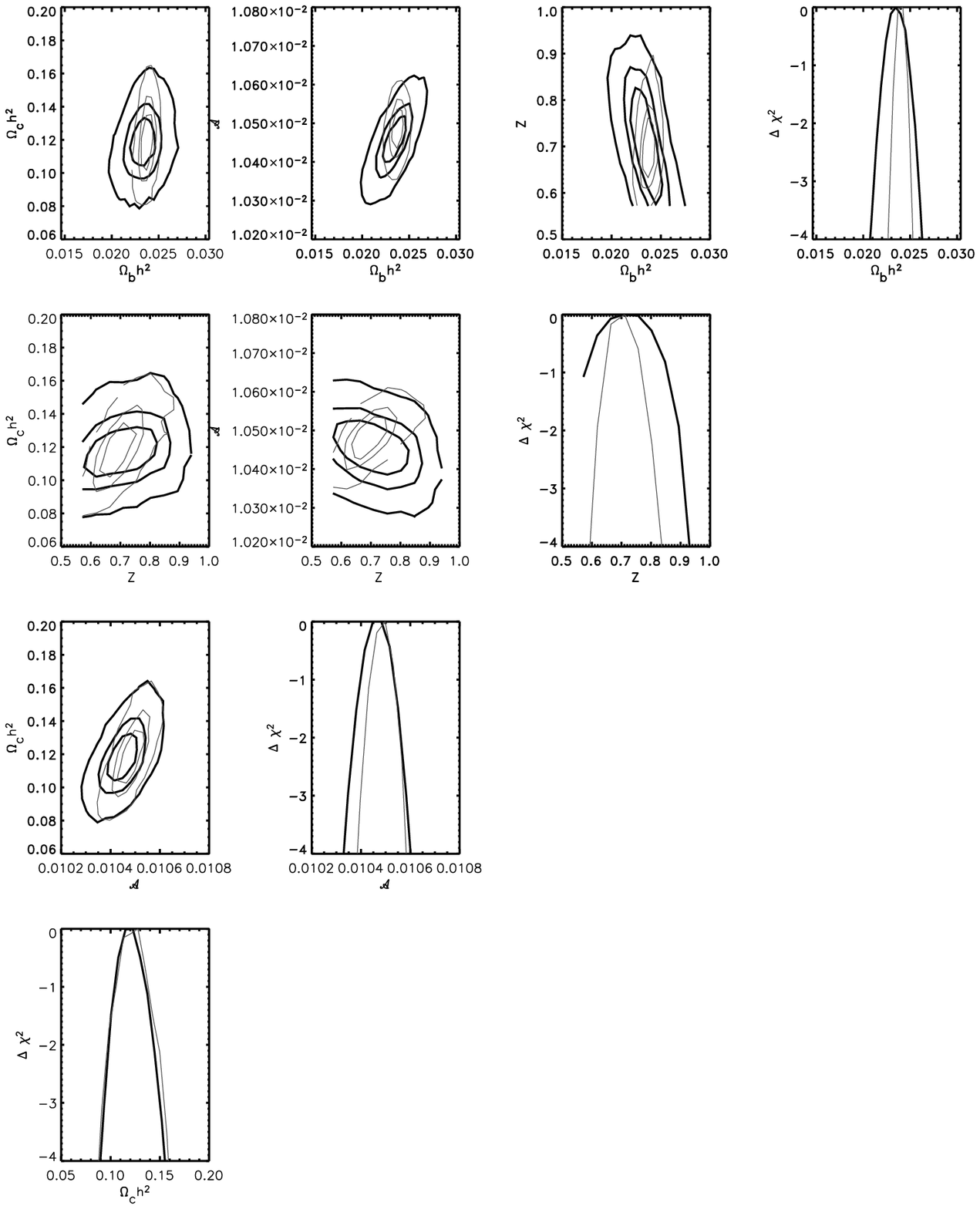}
\caption{Errors on cosmological parameters from a Markov chain run for
a smoothing spline $\pps$ with 5 knots and and a smoothing parameter
of 0.1.  Dark contours denote one sigma marginalized, one sigma joint,
and two sigma joint. Over-plotted with thin, gray contours are the
errors for a $n=1$ flat PPS for comparison. $\Omega_b$ is the baryon
fraction, $\Omega_c$ is the cold dark matter density, ${\cal A}$ is the
present angular size of the sound horizon at the last scattering
surface, and $Z=e^{-2 \tau}$ where $\tau$ is the optical depth.}
\label{fig:cp5.1}
\end{figure*}

Figure (\ref{fig:cp5}) (dark lines) shows the cosmological
parameter likelihood contours for the one sigma marginalized, one
sigma joint, and two sigma joint confidence intervals for the
five-knot spline with no lambda, figure (\ref{fig:cp17}) shows the same
for the seventeen-knot spline with $\a=2\times 10^{-5}$, and figure
(\ref{fig:cp5.1}) shows the same for the five-knot spline with
$\a=0.1$ .  For comparison, the thin gray lines in each figure show
the likelihood contours when the PPS is forced to be
Harrison-Zeldovich.  

We find that the best-fit values for all parameters except $Z$ do not
significantly change as the PPS is allowed more freedom compared to
the scale-invariant case.
The error bars increase, however, especially in the 17-knots case.

Table (\ref{table:CPs}) shows the values and errors for the
cosmological parameters for our two smoothing spline reconstructions
compared to the WMAP errors \cite{Spergel03} and the minimal model
($n=1$) errors.  
The parameter that shows least sensitivity to the increased
flexibility in the PPS is $\Omega_m h^2$, followed by ${\cal A}$. These
parameters are the least degenerate with smooth features in the
PPS. The increased freedom in the PPS increases the errors
significantly (by a factor $\sim 5$ for the five-knot case compared to
the Harrison-Zeldovich case) for $\Omega_b h^2$.  Drastic changes are
also observed in the likelihood contours for $Z$. Following the WMAP
team \cite{Spergel03}, we have imposed a hard prior on $\tau$ of
$\tau<0.3$. While the effect of this prior is not very important for
the scale-invariant case, it sharply cuts the likelihood contours in
the spline cases. As additional freedom in the PPS is allowed, the
recovered best-fit value for $\tau$ is pushed towards the limit
imposed by the prior.
With only the TT and TE power spectra, the optical depth is
significantly degenerate with the PPS shape. We expect the
addition of EE polarization power spectrum will break this degeneracy.

\section{Conclusions}

Cosmological data now offer the ability to go beyond finding preferred
parameters under a given model, and to compare the success and
statistical preference for different models.  Here we have used a
nonparametric (or minimally parametric) method to test a subset of
smooth PPS models.  We have presented smoothing spline reconstructions
of the PPS using CMB data, complementary to other nonparametric
reconstructions presented in the literature.  We applied this
technique to WMAP first-year data.

The smoothing spline reconstruction gives the PPS significantly more
freedom than a power-law or a running spectral index model, and is
poised to recover smooth features in the PPS.  We find that the spline is
not sensitive to small, sharp features in the PPS, but if a feature
such as a running of the spectral index $\alpha_s \gtrsim -0.05$ is present
in WMAP first-year data, the technique can recover it.

We first reconstructed the PPS without any roughness penalty with a
five-knot spline, thus leaving the PPS four extra degrees of freedom
compared to a scale-invariant $n=1$ model.  Then, to quantitatively
assess if the data require extra effective degrees of freedom in the
PPS, we used three different statistical tests.  We first applied cross
validation, the standard statistical method for choosing a smoothing
spline's smoothing parameter.  We found that CV prefers a high
smoothing parameter, indicating a preference for a flat
(scale-invariant) PPS.  We
then compared this result with the AIC and BIC, information criteria
for measuring the statistical significance of additional parameters in
a model.  Both the AIC and BIC strongly indicate no preference for the
extra freedom of the smoothing spline.

Thus, we found no statistically significant indications for any
deviations from a scale invariant ($n=1$) PPS. 
We find that most cosmological parameters, especially the matter density $\Omega_m$,
remain fairly robust to flexibility in the PPS.  However, the
constraints on $\Omega_b h^2$ and to a greater extent 
on the optical depth $\tau$ are significantly weakened. We expect that the
addition of EE-polarization data will break the degeneracy between
optical depth and the PPS shape.
Large-scale structure data, which now overlaps the scales of the CMB
data, could help even more in constraining the cosmological parameters
when the PPS is flexible.  Here we have shown that even using CMB data
alone, the cosmological parameters are quite robust.  Our results are
consistent with \cite{MukWang03c} and \cite{Bridle03}.  In particular,
we find roughly the same increases in errors and changes in the values
of the cosmological parameters as Mukherjee and Wang's \cite{MukWang03c} PPS
reconstructions using 11 wavelet band-powers and top-hat bins and CMB
data only.  The addition of large-scale structure
data naturally shrinks the errors slightly.
 
The smoothing spline is a statistically rigorous tool, and has a solid
Bayesian justification when applied to direct measurements of the
sought-after function 
\cite{Wahba90,GreenSilverman}.  We had to modify the smoothing spline method
due to the complicated transfer function between the CMB data and the
PPS.  With more computing power or faster methods to compute the
 theory $C_\ell$  (such as CMBwarp \cite{cmbwarp}), we could, in theory, include as
many knots as data points, and perform one-fold CV to choose the
optimum smoothing parameter.  Such a reconstruction would be able to
soundly detect features as fine as the data could possibly
distinguish.

\begin{acknowledgments}
We thank Hiranya Peiris and Carlos Hernandez-Monteagudo for invaluable
help.  This work was started after a conversation with John Rice on
non-parametric inference.  We thank Robert Krafty and Ben Wandelt for
discussions.  CS is founded in part by NSF grant NSF0206231, LV is
funded by NASA grant ADP-03-0000-0092, RJ is funded in part by NSF
grant NSF0206231.  We acknowledge the use of the HEALPix package
\cite{healpix} and the use of the Legacy Archive for Microwave
Background Data Analysis (LAMBDA, {\tt
http://lambda.gsfc.nasa.gov}. Support for LAMBDA is provided by the
NASA Office of Space Science.
\end{acknowledgments}

\end{document}